\documentclass[twoside,twocolumn,9pt]{article}
\usepackage{extsizes}

\usepackage[super,sort&compress,comma]{natbib}
\usepackage[version=4]{mhchem}
\usepackage[left=1.5cm, right=1.5cm, top=1.785cm, bottom=2.0cm]{geometry}
\usepackage{balance}
\usepackage{mathptmx}
\usepackage{sectsty}
\usepackage{graphicx} 
\usepackage{lastpage}
\usepackage[format=plain,justification=justified,singlelinecheck=false,font={stretch=1.125,small,sf},labelfont=bf,labelsep=space]{caption}
\usepackage{float}
\usepackage{fancyhdr}
\usepackage{fnpos}
\usepackage[english]{babel}
\addto{\captionsenglish}{%
  
}
\usepackage{array}
\usepackage{droidsans}
\usepackage{charter}
\usepackage[T1]{fontenc}
\usepackage[usenames,dvipsnames]{xcolor}
\usepackage{setspace}
\usepackage[compact]{titlesec}
\usepackage{hyperref}
\usepackage{pdfpages}

\usepackage{epstopdf}

\definecolor{cream}{RGB}{222,217,201}

\usepackage{booktabs}       
\usepackage{subcaption}
\usepackage[list-units=single,detect-all]{siunitx}
\DeclareSIUnit{\calorie}{cal}
\DeclareSIUnit{\kcal}{\kilo\calorie\per\mol}
\DeclareSIUnit{\angstrom}{\text {Å}}
\DeclareSIUnit{\kjoule}{\kJ\per\mol}

\begin{document}

\makeFNbottom
\makeatletter
\renewcommand\LARGE{\@setfontsize\LARGE{15pt}{17}}
\renewcommand\Large{\@setfontsize\Large{12pt}{14}}
\renewcommand\large{\@setfontsize\large{10pt}{12}}
\renewcommand\footnotesize{\@setfontsize\footnotesize{7pt}{10}}
\makeatother

\renewcommand{\thefootnote}{\fnsymbol{footnote}}
\renewcommand\footnoterule{\vspace*{1pt}%
\color{cream}\hrule width 3.5in height 0.4pt \color{black}\vspace*{5pt}} 
\setcounter{secnumdepth}{5}

\makeatletter 
\renewcommand\@biblabel[1]{#1}            
\renewcommand\@makefntext[1]%
{\noindent\makebox[0pt][r]{\@thefnmark\,}#1}
\makeatother 
\renewcommand{\figurename}{\small{Fig.}~}
\sectionfont{\sffamily\Large}
\subsectionfont{\normalsize}
\subsubsectionfont{\bf}
\setstretch{1.125} 
\setlength{\skip\footins}{0.8cm}
\setlength{\footnotesep}{0.25cm}
\setlength{\jot}{10pt}
\titlespacing*{\section}{0pt}{4pt}{4pt}
\titlespacing*{\subsection}{0pt}{15pt}{1pt}

\newcommand{\etal}{\textit{et al.}}

\twocolumn[
  \begin{@twocolumnfalse}

\begin{center}
\LARGE{\textbf{Coordination and thermodynamic properties of aqueous protactinium(V) by first-principle calculations$^\dag$}}
\end{center}

\vspace{0.3cm}

\begin{center}
\large{Hanna Oher,\textit{$^{a,b}$} Jeremy Delafoulhouze,\textit{$^{c}$} Eric Renault,\textit{$^{c}$} Valérie Vallet,\textit{$^{d}$} and Rémi Maurice$^{\ast}$\textit{$^{a,b}$}}
\end{center}

\vspace{0.3cm}

\noindent\normalsize{Protactinium ($Z$ = 91) is a very rare actinide with peculiar physico-chemical properties. Indeed, although one may naively think that it behaves similarly to either thorium or uranium by its position in the periodic table, it may in fact follow its own rules. Because of the quite small energy gap between its valence shells (in particular the 5$f$ and 6$d$ ones) and also the strong influence of relativistic effects on its properties, it is actually a challenging element for theoretical chemists. In this article, we combine experimental information, chemical arguments and standard first-principle calculations, complemented by implicit and explicit solvation, to revisit the stepwise complexation of aqueous protactinium(V) with sulfate and oxalate dianionic ligands (\ce{SO4^{2-}} and \ce{C2O4^{2-}}, respectively). From a methodological viewpoint, we notably conclude that it is necessary to at least saturate the coordination sphere of protactinium(V) to reach converged equilibrium constant values. Furthermore, in the case of single complexations (\textit{i.e.} with one sulfate or oxalate ligand bound in the bidentate fashion), we show that it is necessary to maintain the coordination of one hydroxyl group, present in the supposed \ce{[PaO(OH)]^{2+}} precursor, to obtain coherent complexation constants. Therefore, we predict that this hydroxyl group is maintained in the formation of 1:1 complexes while we confirm that it is withdrawn when coordinating three sulfate or oxalate ligands. Finally, we stress that this work is a first step toward the future use of theoretical predictions to elucidate the enigmatic chemistry of protactinium in solution.} 

 \end{@twocolumnfalse} \vspace{0.3cm}

\vspace{0.6cm}

]

\footnotetext{\textit{$^{a}$~Subatech, UMR CNRS 6457, IN2P3/IMT Atlantique/Universit\'e de Nantes, 4 rue A. Kastler, 44307 Nantes Cedex 3, France}}
\footnotetext{\textit{$^{b}$~Univ Rennes, CNRS, ISCR (Institut des Sciences Chimiques de Rennes) -- UMR 6226, F-35000 Rennes, France; E-mail: remi.maurice@univ-rennes.fr}}
\footnotetext{\textit{$^{c}$~Nantes Université, CNRS, CEISAM UMR 6230, F-44000 Nantes, France}}
\footnotetext{\textit{$^{d}$~Univ. Lille, CNRS, UMR 8523-PhLAM-Physique des Lasers, Atomes et Molécules, F-59000 Lille, France}}
\footnotetext{\dag~Electronic Supplementary Information (ESI) available: Preparatory study on molecular geometries, which includes comparisons of methods, basis sets and codes. }

\section{Introduction}

Protactinium ($Z =91$) is an enigmatic element for the chemists\cite{Wilson:2012}. While it is located in between thorium and uranium in the periodic table, it may display unique properties that severely hinder experimental studies and interpretations. For instance, the five valence electrons of its free atom may spread in numerous ways on the valence 5$f$, 6$d$, 7$s$ and 7$p$ shells, which generates a quantum chaos of states even well before ionization\cite{Viatkina:2017, Naubereit:2018a, Naubereit:2018b}. In solution, protactinium chemistry is expected to be dominated by its +V oxidation state (formally free of valence electron), with potential occurrence under specific reducing conditions of the unstable +IV one (displaying one unpaired 5$f$ electron in the ground state) \cite{Marquardt:2004}. Therefore, the aqueous chemistry of protactinium could well have been dominated by easily identifiable and computable species.

Unfortunately, Pa(V) and Pa(IV) are prone to hydrolysis, polymerization and precipitation and other issues such as a strong tendency to adsorb into glass, \textit{etc.} \cite{LeNaour:2022}. Furthermore, the intrinsic radiations of its isotopes imply specific safety regulations to be followed. Consequently, the speciation of Pa(V) and Pa(IV) in aqueous solution is largely unknown, which further complicates the performance of a theoretical study dedicated to the coordination and thermodynamic properties of these; thus, one has to rely on \textit{ad hoc} research hypotheses.

In this work, we will focus on protactinium(V) since it is expected to have a closed-shell electronic ground state in most of its associated chemical species. Therefore, single-reference approaches such as Kohn-Sham density functional theory (DFT) or second-order M{\o}ller-Plesset perturbation theory (MP2) may be \textit{a priori} safely used, contrary to Pa(IV) systems. Despite this apparent simplicity in the electronic structure theory part, challenges remain: solvation and/or hydrolysis issues must be tackled and relativistic effects, especially the scalar ones, should be accounted for. In fact, since little experimental data regarding speciation and complexation constants available, the choice of the Pa(V) chemical complexes will be severely restrained. Anyway, we will show that it is necessary to combine experimental information, chemical arguments and outcomes of calculations to reach conclusions.

The paper is organized as follows. First, the choice of the systems and our working hypotheses will be explained. Then, computational details will be given, prior to results and discussion, organized by increasing solvation and/or chemical complexity.

\section{Choice of the systems and working hypotheses}

The chemistry of Pa(V) is characterized by a general lack of experimental and/or computational data \cite{le2019protactinium}. In this work, we aim at showing that standard quantum chemical calculations may be of help to predict the relative stabilities of Pa(V) complexes, based on comparisons between theoretical and experimental data. Therefore, it is necessary to make a quick survey of known experimental data to justify our choice of the systems under study. In particular, the available experimental data should ideally help us to formulate relevant hypotheses concerning the nature of the complexes that are involved in the equilibrium processes at play. After this, we will introduce additional working hypotheses that are required to define a practical computational strategy, systematically improvable in terms of solvation and/or chemical models.

The identification of Pa(V) species is a difficult task. Several types of complementary experiments may be performed, none of them being independently conclusive. We may classify them in two main types, (i) experiments at tracer scale and (ii) spectroscopy experiments (ponderable scale). In (i), we may encounter methods to determine the molecular charge (ion chromatography and/or migration experiments under an electric field such as capillary electrophoresis) and also methods to determine equilibrium constants, based on competition between two distinct phases (\textit{e.g.} liquid-liquid extraction). In (ii), we may find X-ray absorption data for instance, based on near-edge or extended structures (XANES and EXAFS, respectively). By taking profit of the X-ray absorption spectrum dependence on the molecular structure, especially in the EXAFS regime, one may refine guessed structural models by a fitting procedure, leading in the end to an experimental structure.

It is important to understand that equilibrium constants are determined by fitting procedures, implying that an underlying thermodynamic model is selected. Because the activity of water in water is set to 1, all the reactions are defined up to a given number of water molecules. This is a strong limitation of the liquid-liquid extraction approach. For instance, Mendes \textit{et al.} \cite{mendes2013thermodynamic} have shown that Pa(V) can form a 1:1 complex with DTPA (diethylenetriaminepentaacetic acid) and have determined the associated formation constant for the following reaction:
\begin{equation}
    \ce{PaO(OH)^{2+} + DTPA^{5-} + 3H+ <=>[$\beta_{app}$]  Pa^{V}-DTPA + x H2O}
\end{equation}
\noindent where \ce{Pa^{V}-DTPA} is either \ce{PaO(H_2DTPA)} or \ce{Pa(DTPA)} and $x$ = 1 or $x$ = 2, respectively. To make a long story short, the neutral charge was determined by capillary electrophoresis, and the potentially formed complexes justified by chemical arguments and quantum mechanical calculations. Clearly, the absence of X-ray absorption data rules out this system for a first computational study on equilibrium constants, since we are stuck with an ambiguity in the nature of the formed complex.

EXAFS data of Pa(V) have been acquired in fluoric acid (\ce{HF}), sulfuric acid (\ce{H_2SO_4}) and oxalic acid (\ce{H_2C_2O_4}) concentrated media \cite{le2005first,mendes2010thermodynamical,de2014exafs}. Concentrated media are interesting because we may expect to saturate the first coordination sphere of the Pa(V) ion and end up with a simpler speciation than at intermediate concentrations, \textit{i.e.} we may observe single well-defined chemical species. While in fluoric acid the mono-oxo bond of the \ce{[PaO(OH)]^{2+}} starting species is cleaved (up to eight fluorides being coordinated \cite{de2014exafs,wilson2016structural}), it is observed in both the sulfuric and oxalic media. Therefore, we may expect similar complexes to be formed. Since in the oxalic acid case, the analysis of the data led to the firm identification of the \ce{[PaO(C_2O_4)_3]^{3-}} complex, we thus hypothesize the formation of the \ce{[PaO(SO_4)_3]^{3-}} complex in sulfuric acid medium. We recall here that our objective is to determine relative complexation constants, meaning that it is important to compare systems of similar natures to obtain accurate values by error cancellations, as was done in the past for exploring the astatine ($Z$ = 85) basic chemistry \cite{Champion:2011, Champion:2013, sergentu2016scrutinizing, sergentu2016advances, guo2016heaviest}.

Having suspected that complexes of similar natures may be formed at high \ce{H_2SO_4} and \ce{H_2C_2O_4} concentrations, in particular of the \ce{[PaOX_3]^{3-}} type, \ce{X} being either \ce{(SO_4)} or \ce{(C_2O_4)}, we continue our survey with the corresponding liquid-liquid extraction data\cite{le2019protactinium,le2005first,mendes2010thermodynamical,di2009complex}. In these studies, the formations of 1:1, 1:2 and 1:3 complexes of Pa(V) with sulfate or oxalate ligands were hypothesized and described by the following chemical equilibria:
\begin{equation}
    \ce{PaO(OH)^{2+} + i X^{2-} + H+ <=>[$\beta_{app,i}$] [PaO(X)_i]^{3-2i} + H2O}
\end{equation}
\noindent where \ce{\beta_{app,i}} are the (global) apparent formation constants and $i$ varies from 1 to 3. Again, since the formed species are defined up to a given number of water molecules, only apparent constants have been determined so far. By data analysis, the formation of the 1:1 to 1:3 complexes were confirmed in both the cases and the apparent formation constants have been determined at \SI{25}{\celsius} by a fitting procedure~\cite{di2009complex,mendes2010thermodynamical} (see Table~\ref{tab:log_beta_exp}). 

\begin{table}[h]
\small
  \caption{\ Experimental apparent formation constants of Pa(V) complexes determined at \SI{25}{\celsius} and deduced ligand-exchange experimental equilibrium constants.}
  \label{tab:log_beta_exp}
  \begin{tabular*}{0.48\textwidth}{@{\extracolsep{\fill}}llll}
    \hline
     \SI{25}{\celsius} &  \ce{SO4^{2-}}~\cite{di2009complex} & \ce{C2O4^{2-}}~\cite{mendes2010thermodynamical} & log $K_{exc,i}^{exp}$ \\
    \hline
    log $\beta_{app,1}$ & \num{3.9\pm 0.2} & \num{7.4\pm 0.3} & \num{3.5\pm 0.4}\\
    log $\beta_{app,2}$ & \num{7.0\pm 0.2} & \num{14.5\pm 0.8} & \num{7.5\pm 0.9}\\
    log $\beta_{app,3}$ & \num{8.6\pm 0.2} & \num{19.1\pm 0.7} & \num{10.5\pm 0.8}\\
    \hline
  \end{tabular*}
\end{table}

From the log $\beta_{app,i}$ values reported in Table~\ref{tab:log_beta_exp}, we observe a similar trend in both the series, with the log $\beta_{app,2}$ values being nearly twice the log $\beta_{app,1}$ ones, and the log $\beta_{app,3}$ ones being nearly 2.5 times larger than the log $\beta_{app,1}$ ones. Because of this, we hypothesize that for each $i$ value, the two products, \textit{i.e.} \ce{[PaO(SO4)_i]^{3-2i}} and \ce{[PaO(C2O4)_i]^{3-2i}}, have similar natures. Therefore, even if the nature of the ending complexes is not (yet) fully known, we propose to focus on three ligand-exchange reactions:
\begin{equation}\label{eq:reaction1}
\ce{[PaO(SO_4)_i]^{3-2i} + i C_2O_4^{2-} <=>[$K_{exc,i}$] [PaO(C_2O_4)_i]^{3-2i} + i SO_4^{2-}}
\end{equation}
\noindent The determination of the experimental exchange constants is in this case trivial as it derives from the difference of the individual apparent complexation constants:
\begin{equation}
\text{log}~K_{exc,i}^{exp} =  \text{log}~\beta_{app,i}~[\ce{C2O4^{2-}}] - \text{log}~\beta_{app,i}~[\ce{SO4^{2-}}]
\end{equation}
\noindent It is important to recall that the mono-oxo bond has been observed in the \ce{[PaO(C_2O_4)_3]^{3-}}, \textit{i.e.} for the complex of highest ligation with the ligand that best bind to Pa(V). As a consequence, we assume that the mono-oxo bond is maintained in all the complexes of interest. Thus, the main point to be addresses relates to the potential departure of the hydroxyl group, present in the starting \ce{[PaO(OH)]^{2+}} species, as mentioned earlier.

\begin{figure}[h!]
\centering
  \includegraphics[height=6cm]{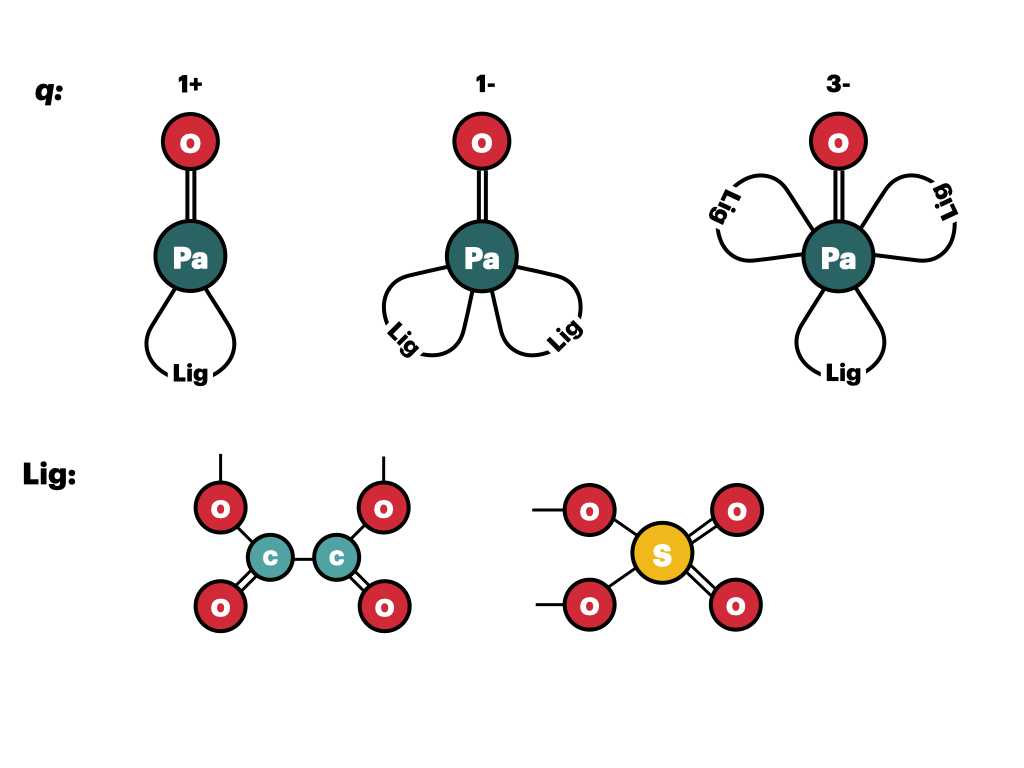}
  \caption{Schematic representation of the geometrical models for the Pa(V) complexes that are used to compute the $\text{log}~K_{exc,i}$ values.}
  \label{fig:model_systems}
\end{figure}

Our computational study aims at fairly reproducing the $\text{log}~K_{exc,i}^{exp}$ values by using solvation and chemical models of increasing complexities. To achieve this goal, we formulate the following working hypotheses:

\begin{enumerate}
    \item In all the complexes, the ligands bind in a bidentate fashion (see Figure~\ref{fig:model_systems});
    \item We can compute the log of the exchange constants by means of a thermodynamic cycle\cite{ho2015thermodynamic}, with computation of the log of the gas phase constants at our reference level of theory and computation of the solvation energies of the species with a polarizable continuum model (PCM);
    \item We assume that the errors done by our PCM approach concerning the solvation of the free \ce{SO4^{2-}} and \ce{C2O4^{2-}} ligands are similar, meaning that explicit solvation will not be applied on the free ligands (note that the opposite would potentially turn out to be a dead end owing the complexity of the generated microsolvated structures\cite{Gao:2004, Gao:2005}).
\end{enumerate}

\noindent Also, we recall the previously formulated hypotheses still apply, meaning that we assume that the mono-oxo bond is present in all the complexes and also that for each $i$ reaction, the corresponding sulfate and oxalate complexes have a similar nature.

Finally, we define the log of the exchange constants as follows ($T$ is fixed at 298.15 K and $P$ at 1 atm):
\begin{equation}
\text{log}~K_{exc,i} = -\frac{\Delta G_{exc,i}^0}{2.303 R T } 
\end{equation}

\section{Computation details}

Geometries and electronic structures of Pa(V) are expected to be well described by single-reference approaches, due to the closed-shell nature of the valence-free reference \ce{Pa^{5+}} ion. We have thus directly restricted our preparatory study to the MP2 and Kohn-Sham DFT methods, with inclusion of scalar relativistic effects (via the Douglas-Kroll-Hess Hamiltonian\cite{douglas1974quantum,hess1986relativistic,Jansen:1989} or a suited relativistic pseudopotential), and in the latter case the use of a hybrid exchange-correlation functional. We have actually compared results obtained with the MP2\cite{Moller:1934}, DFT/B3LYP\cite{becke1998new} and DFT/PBE0\cite{Perdew1996, adamo1999} methods, with different basis sets (with an all-electron basis \cite{pantazis2011a} \textit{vs.} bases \cite{Kendall:1992,ecp-Cao-JMST2004-673-203,weigend2005balanced-Clbasisset} combined with a relativistic pseudopotential for Pa\cite{ecp-Cao-JCP2003-118-487}), and different codes (Gaussian\cite{g16}, ORCA 4.2.1~\cite{neese2017orca} and OpenMolcas\cite{aquilante2020modern}). The results and a more complete discussion can be found in Supporting Information. Our main findings are:

\begin{enumerate}
    \item Basis sets combined with pseudopotentials generally give geometries in good agreement with the employed all-electron basis, unless a specific and unrecommended implementation is used.
    \item The use of the def2-TZVP and aug-cc-pVTZ basis sets on the light atoms (\textit{i.e.} all atoms besides Pa) leads to a significant improvement over the previous results that were based on the former 6--31+G* ones \cite{mendes2010thermodynamical}.
    \item All the tested approaches are in good agreement with the EXAFS structure, with perhaps a slight advantage of the DFT/PBE0 approach, essentially concerning the mono-oxo bond distance.
\end{enumerate}

In the remainder of the article, the gas-phase geometries, harmonic frequencies, and gas-phase free energies of the oxalate and sulfate complexes with Pa(V) were obtained using the Gaussian~16.A.03 software~\cite{g16}. All the reported structures correspond to minima of the energy (no negative frequency). The Kohn-Sham equation was solved by using the hybrid PBE0 exchange-correlation functional~\cite{Perdew1996, adamo1999}. In these calculations, the def2-TZVP (second generation of triple-$\zeta$ plus polarization quality) basis set~\cite{weigend2005balanced-Clbasisset} has been used for all the light elements (H, C, O, S). For the protactinium atom, the small-core scalar relativistic effective core potential ECP60MWB was used~\cite{ecp-Cao-JCP2003-118-487} along with its associated basis set, formed by contracting a 14s13p10d8f set into a 10s9p5d4f one~\cite{ecp-Cao-JMST2004-673-203}.

The aqueous solvation free energies were computed at the Hartree-Fock (HF) level with the same basis sets and pseudopotentials as used in the DFT/PBE0 calculations. To mimic the long-range solvent effects, the conductor-like polarizable continuum model (CPCM)~\cite{Barone:1998aa-cpcm, cossi2003cpcm} was used as it is implemented in the Gaussian 16 program package. In the absence of any alternative, the UFF radius was for the Pa atom (\SI{1.712}{\angstrom}), while the rest of the cavity was based on UAHF radii\cite{Barone:1997} for the other atoms/groups (\ce{O_{oxo}}=\SI{1.59}{\angstrom}, \ce{O_{ligand}}=\SI{1.35}{\angstrom}, \ce{C}=\SI{1.68}{\angstrom}, \ce{S}=\SI{1.98}{\angstrom}, \ce{OH-}=\SI{1.59}{\angstrom}, \ce{H2O}=\SI{1.68}{\angstrom}). We recall here that the UAHF radii depend on the charge and environment of the atom of interest. For the bidentate \ce{SO4^{2-}} and \ce{C2O4^{2-}} ligands, we have considered \textit{ad hoc} equal $-$1/2 charges on the oxygen atoms, hence the \ce{O_{ligand}} radius value. Also, note that no sphere is created around the H atoms (united-atom approach), which explains the \ce{OH-} and \ce{H2O} notations. Since those radii were initially optimized with the \textit{alpha} scaling factor of 1.2, it was manually set to this value in our calculations. 

Choice was made to apply the UAHF model instead of the UAKS after initial calculations concerning the geometry of the \ce{[PaO(C2O4)3]^{3-}} complex and the determination of the $\text{log}~K_{exc,3}$ value. In fact, a positive solvation contribution to $\text{log}~K_{exc,3}$ was obtained with the UAKS model, resulting in an unphysical negative $\text{log}~K_{exc,3}$ value. In a previous study related to astatine chemistry\cite{sergentu2016scrutinizing}, it was also found that the UAHF model was more accurate than the UAKS one to compute ligand-exchange reaction constants, the UAKS one being more accurate only if applied in a single-point fashion at the UAHF geometries. In view of finding a compromise between accuracy and simplicity of application (no need for extra single-point calculations), we have thus here retained the UAHF model for the production calculations.

\section{Results and discussion}

In this section, we will introduce models of increasing complexity. First, we will look at the gas-phase ligand-exchange reaction constants, and complement these by an implicit solvation model. Then, we will also include explicitly water molecules in the quantum chemical framework, meaning that we will use a combined implicit and explicit solvation model. Finally, we will discuss the possibility for maintaining the hydroxyl group on both the left-hand and right-hand sides  of the ligand-exchange reactions. At each stage, we will compare the computed $\text{log}~K_{exc,i}$ values to the $\text{log}~K_{exc,i}^{exp}$ ones.

\subsection{Gas phase (GP) and implicit solvation (PCM)}

The zeroth-order description of solvation consists in fully neglecting any solvent effect, implying that gas-phase reaction constants are actually first sought. In the process, the geometries of the sulfate and oxalate complexes of interest are determined, together with the structures of the free ligands. For all the complexes, we find that binding in bidentate fashion is by far preferable (which justifies our previous hypothesis number \#1).

As already mentioned, the EXAFS structure of the \ce{[PaO(C2O4)3]^{3-}} complex has been reported by Mendes~{\etal}\cite{mendes2010thermodynamical}. We note that attempts to determine the structure of the analogous \ce{[PaO(SO4)3]^{3-}} complex have been reported\cite{le2005first,di2008behaviour}, but that information concerning the binding modes is still lacking, meaning that even if the presence of the short mono-oxo bond was confirmed, it is hard to know how accurate the other Pa--\ce{O_{ligand}} bond distances are. Therefore, we only retain the EXAFS structure of the \ce{[PaO(C2O4)3]^{3-}} complex for comparison purposes.

As shown in Supporting Information, the DFT/PBE0 approach leads to a good agreement with the experimental structure, despite a moderate discrepancy on the Pa--\ce{O_{oxo}} bond distance (\SI{1.86}{\angstrom} \textit{vs.}\SI{1.75}{\angstrom} in the experiment). At this stage, the computational/experimental origin of this discrepancy is not known, and it may relate to experimental and/or computational uncertainty. Addressing this issue would require generating new EXAFS experiments and/or new fits of existing data and also performing an extensive electronic structure theory benchmark, which is out of the scope of the present paper. In fact, we remark that for all the Pa--\ce{O_{ligand}} bond distances of interest (\textit{i.e.} the ones of the bound atoms) as well as for the Pa--\ce{C} bond distances, the mean absolute deviation between the computed and experimental values is below \SI{0.02}{\angstrom}. We thus assume that the retained DFT/PBE0 structure are overall accurate enough for further proceeding with the computation of ligand-exchange reaction constants. Moreover, we remark that a Pa--\ce{O_{oxo}} bond distance of \SI{1.86}{\angstrom} seems significantly longer than the one in bare \ce{PaO^{3+}} (\SI{1.68}{\angstrom}) and in the \ce{PaO(OH)^{2+}} (\SI{1.71}{\angstrom}) precursor. The three studied reactions have been renamed as R1, R2 and R3, depending on the number of ligands involved in the reaction. The gas-phase structures of the complexes involved in these reactions are displayed in Figure S1. Results are reported in Table~\ref{tab:constants} (see more specifically the ``GP'' column). 
 
 \begin{table}[h]
\small
  \caption{\ Computed values of the logarithms of the ligand-exchange reaction constants as functions of the employed solvation model (see text). The experimental values from Table~\ref{tab:log_beta_exp} are repeated for convenience.}
  \label{tab:constants}
  \begin{tabular*}{0.48\textwidth}{@{\extracolsep{\fill}}llllll}
    \hline
    $\text{log}~K_{exc,i}$ & GP &PCM&\multicolumn{2}{c}{PCM + $n$ \ce{H2O}} & Expt.   \\ \cline{4-5}
     &   &&FCS&  FCS+1&    \\
    \hline
    R1  & 6.0 &9.4&-&- & 3.5\\
    R2  & 2.6 &19.9&12.0&8.8& 7.5 \\
    R3  & 2.8 &16.1&16.1&14.6 & 10.5\\
    \hline
  \end{tabular*}
\end{table}

For all the computed species, the obtained lowest-energy structure was separated by at least \SI{10}{\kjoule} with any other conformer or isomer (in terms of Gibbs free energy at \SI{298.15}{\kelvin}). Therefore, we have used only one structure per species to compute the reaction constants. As can be seen in Table~\ref{tab:constants}, the gas-phase ligand-exchange reaction constants do not follow at all the experimental trend, going from $\text{log}~K_{exc,1}$ = 6.0 (R1) to $\text{log}~K_{exc,3}$ = 2.8 (R3). This may be a net indication that solvent effects may be at play, suggesting that important solute-solvent interactions must somehow be accounted for~\cite{cossi2003cpcm}. 
 
As a first solvation model, we have included a correction in the free energy of each species arising from a PCM model (see the Computational details section). In computing those solvation-induced corrections to the free energies, it is important to check that the geometry that is optimized with application of the PCM is of the same nature as the gas-phase one. Furthermore, in our calculations, we have employed the UAHF model, meaning that both the gas phase and PCM geometries are obtained at the HF level of theory. Therefore, we not only need to check that those two match well, we also need to check that the gas phase DFT/PBE0 and HF geometries are also in qualitative accord. In all the cases, the GP(DFT/PBE0), GP(HF) and PCM(HF) structures qualitatively matched, meaning that consistent corrections to the free energies of the species of interest were obtained. The deduced values of the logarithms of the reaction constants are listed in Table~\ref{tab:constants} (see in particular the ``PCM'' column). 

Again, even if now $\text{log}~K_{exc,1}$ is the weakest of the three, the computational trend differs drastically from the experimental one. Since we observe important changes by introducing the PCM, we may already conclude that solvation is crucial in this case, even if its treatment remains to be more precise. Within the framework of a static quantum chemical approach, we have two main degrees of freedom, we may (i) explicitly treat some of the solvent (here water) molecules or (ii) change the nature of the chemical species that are involved in the studied equilibria, which will be done in the next subsections. Note that the addition of explicitly treated water molecules in the first coordination sphere of the Pa(V) ions may also lead to better geometrical models for the complexes of interest. 

\subsection{Implicit plus explicit solvation (PCM + $n$ \ce{H2O})}
The combination of implicit and explicit solvation usually improves the quality of thermodynamic results at the price of an increased complexity in the conformational search of the solute geometries~\cite{cossi2003cpcm}. A key point lies in the choice of the number of water molecules. At each addition of one water molecule, one may introduce key interactions such as a coordination or a hydrogen bond. Since we compute ligand-exchange reaction constants, it is also important to ensure that the added interactions on the left-hand and right-hand sides of the reaction are of similar nature\cite{sergentu2016advances}. As a consequence, the computed $\text{log}~K_{exc,i}$ values may not monotonously evolve with the number of explicitly treated water molecules, and it may not be worth reporting each and every step.

At this stage, it is important to come back to our computational hypothesis \#3, according to which error cancellations allow us to avoid performing a tricky microsolvation study of the free oxalate and sulfate ligands (many water molecules being required to fill their first solvation spheres\cite{Gao:2004, Gao:2005}). In fact, two independent PCM studies were successful in reproducing an experimental trend that involves the free oxalate or the sulfate ligand: Aquino \textit{et al.}\cite{Aquino:2000} used a \SI{-986.6}{\kjoule} solvation energy for the free oxalate ligand to study its successive complexation with the \ce{Al^{3+}} ion and Lee and McKee\cite{Lee:2011} recommended a \SI{-1028.8}{\kjoule} solvation energy for the free sulfate ligand to study the dissolution of salts. We note that the resulting difference in solvation energies, \SI{-42.2}{\kjoule}, is pretty close to the \SI{-40.0}{\kjoule} that we have obtained with the UAHF model, which supports our hypothesis. As a result, water molecules are only added to the protactinium complexes. We thus rewrite our generic ligand-exchange reaction as: 
 \begin{equation}\label{eq:reaction2}
 \begin{split}
 \ce{[PaO(SO_4)_i (H2O)_n]^{3-2i} + i C_2O_4^{2-} <=>[$K_{exc,i}$]} \\ \ce{[PaO(C_2O_4)_i(H2O)_n]^{3-2i} + i SO_4^{2-}}    
 \end{split}
 \end{equation}
\noindent where \textit{n} has to be carefully chosen, as mentioned above. 

In practice, a stepwise water addition was performed, with a conformational search based on multiple starting structures, having for instance ligands bound in monodentate or bidentate fashions, and also water molecules interacting with either the Pa(V) ion (coordination bonds) or the ligands (hydrogen bonds). Out of this stepwise study, we concluded that we needed to at least saturate the first coordination sphere of the Pa(V) ion to obtain similar structures for each pair of sulfate and oxalate complexes. Therefore, the intermediate steps with an incomplete first coordination sphere are not reported here.

\subsubsection{First coordination sphere (FCS)} 
\begin{figure*}[!htbp]
 \centering
	\begin{minipage}[t]{0.3\linewidth}	
		\includegraphics[height=3cm]{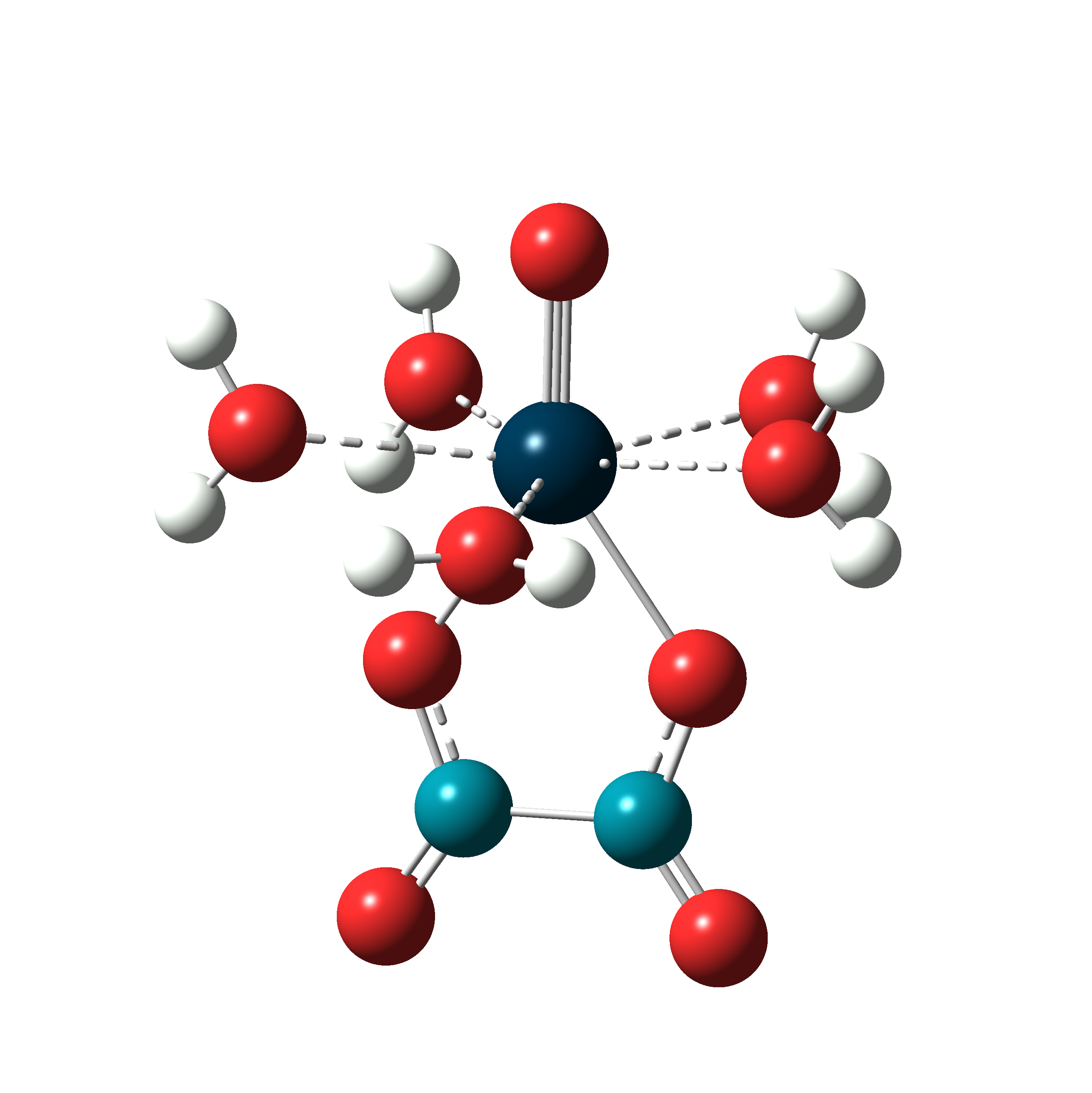}
		\subcaption{\ce{[PaO(C2O4)(H2O)5]^{+}}}
	\end{minipage}
	\begin{minipage}[t]{0.3\linewidth}
		\includegraphics[height=3cm]{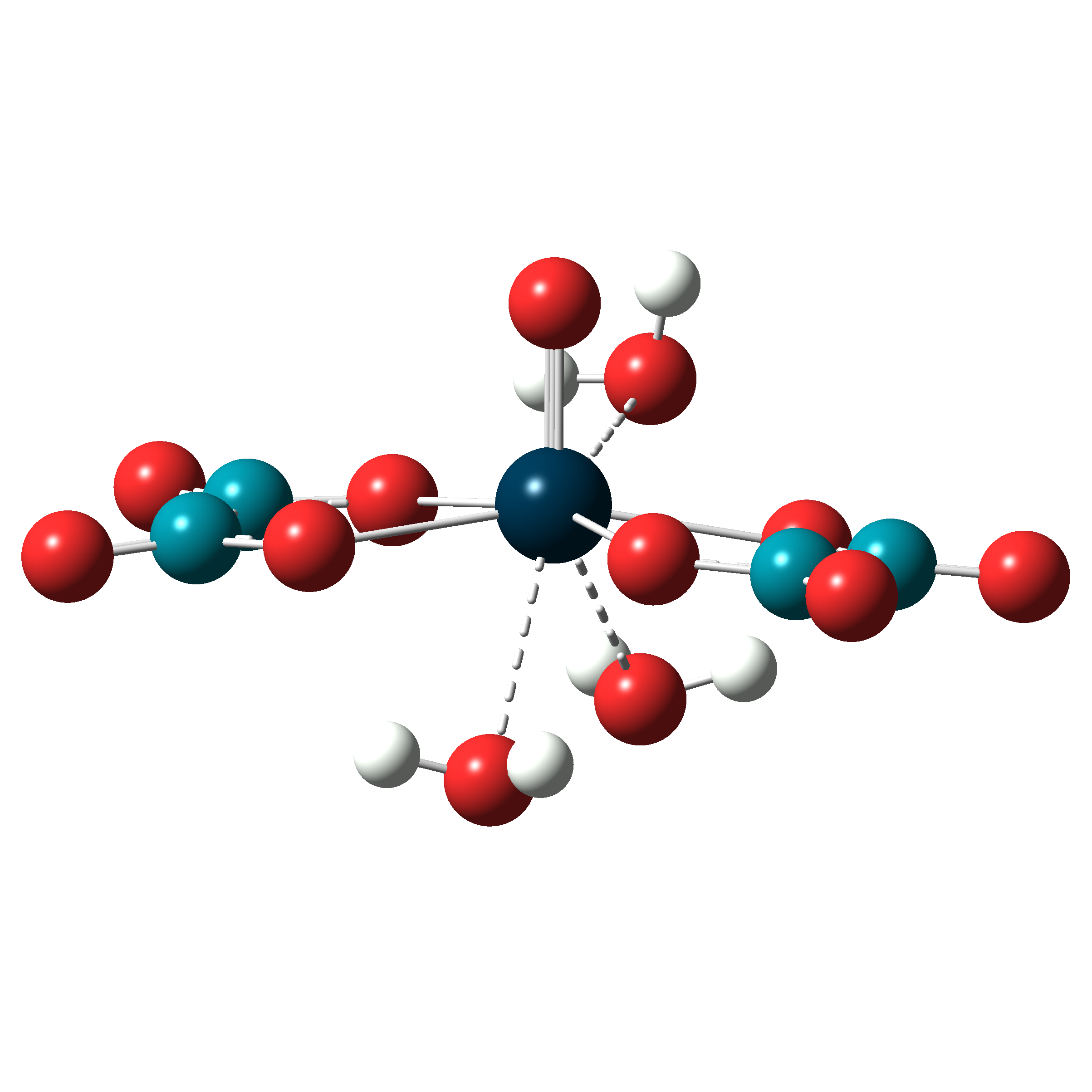}
		\subcaption{\ce{[PaO(C2O4)2(H2O)3]^{-}}}
	\end{minipage}
	\begin{minipage}[t]{0.3\linewidth}
		\includegraphics[height=3cm]{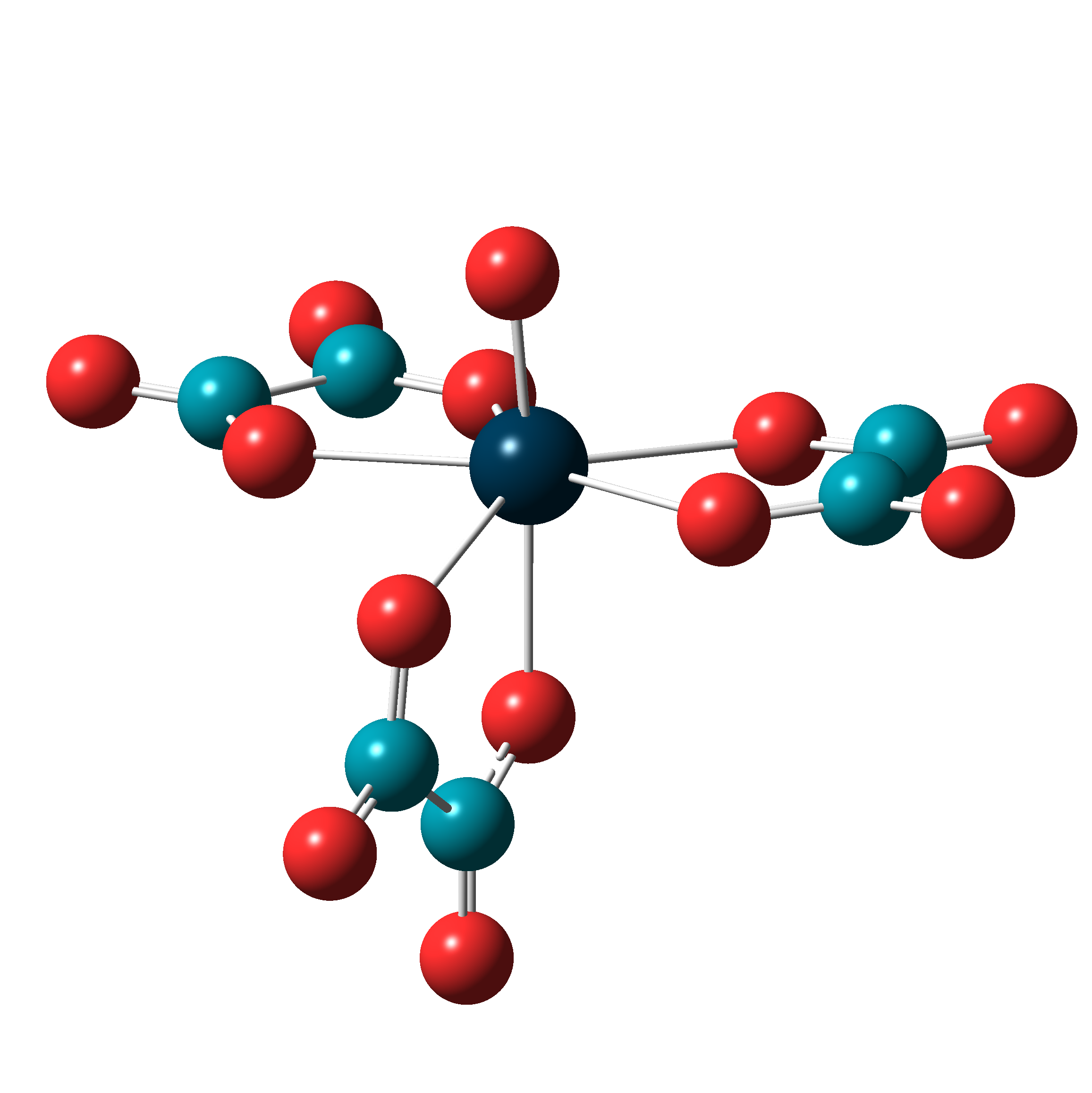}
		\subcaption{\ce{[PaO(C2O4)3]^{3-}}}
	\end{minipage}
	\begin{minipage}[t]{0.3\linewidth}	
	\includegraphics[height=3cm]{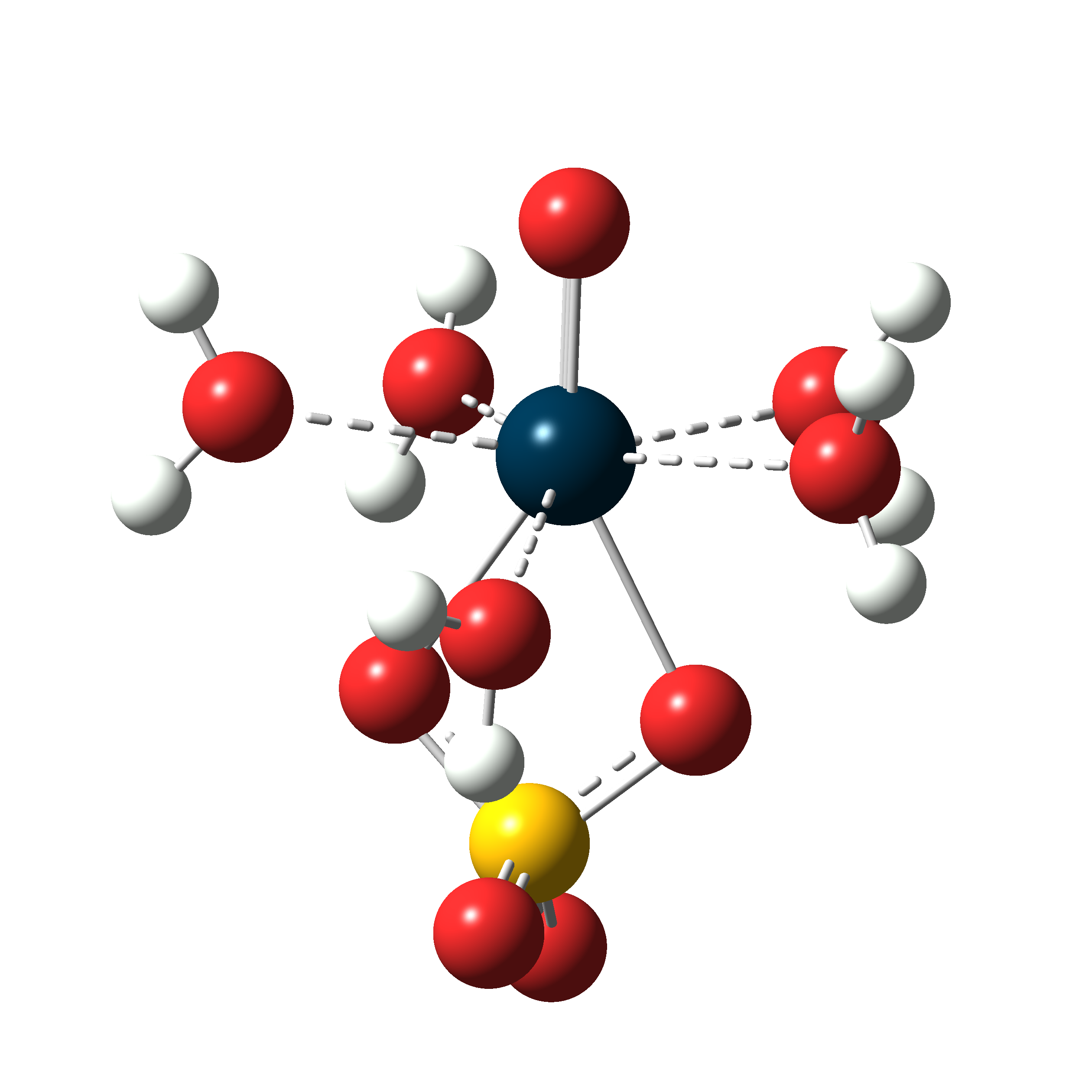}
	\subcaption{\ce{[PaO(SO4)(H2O)5]^{+}}}
\end{minipage}
\begin{minipage}[t]{0.3\linewidth}
	\includegraphics[height=3cm]{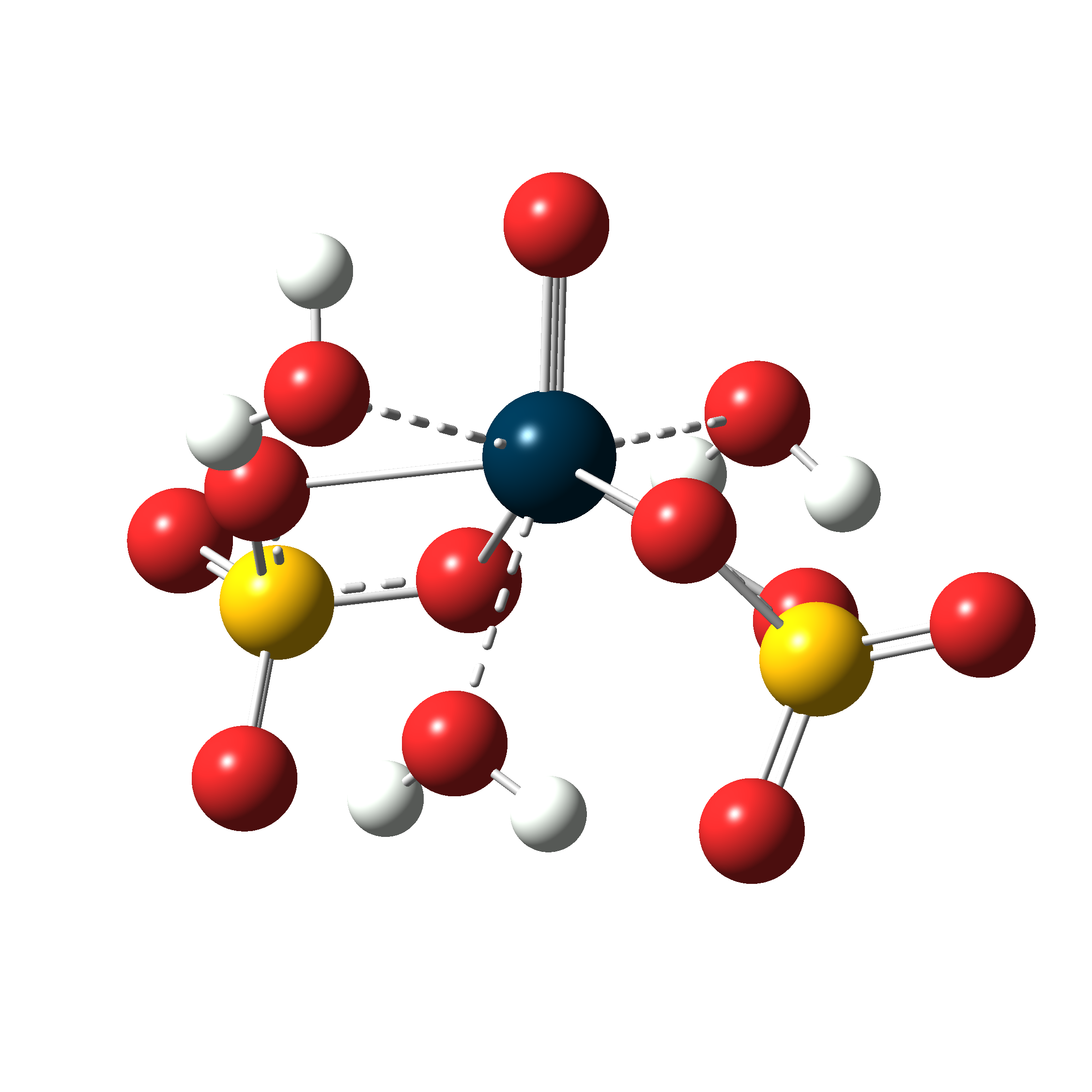}
	\subcaption{\ce{[PaO(SO4)2(H2O)3]^{-}}}
\end{minipage}
\begin{minipage}[t]{0.3\linewidth}
	\includegraphics[height=3cm]{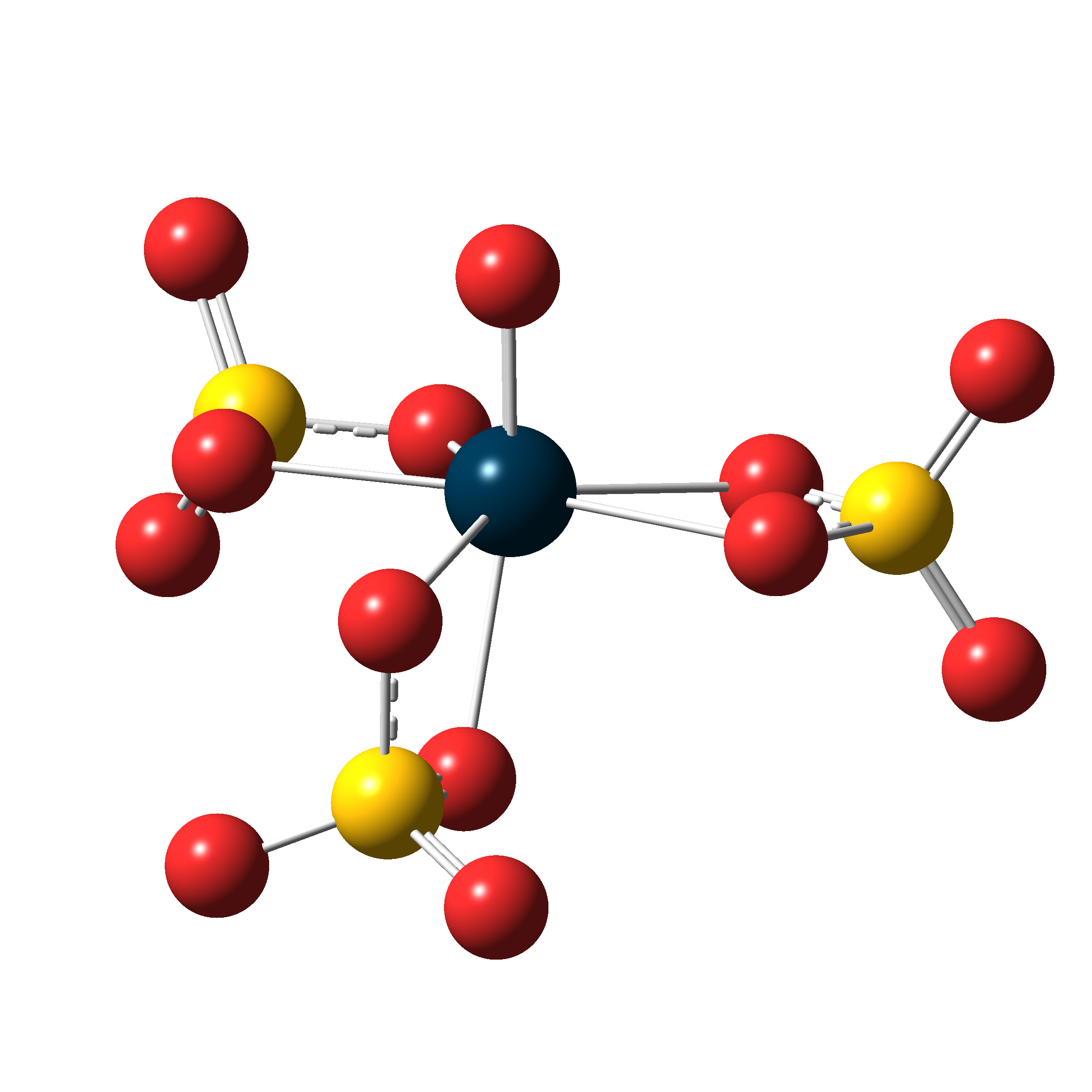}
	\subcaption{\ce{[PaO(SO4)3] ^{3-}}}
\end{minipage}
 \caption{Representations of the GP(DFT/PBE0) structures of the Pa(V) complexes, obtained with just-saturated first-coordination spheres (FCS). Color code: dark blue (Pa), yellow (S), red (O), light blue (C) and white (H).}
 \label{fig:fcs-6structures}
\end{figure*}

The structures of the Pa(V) complexes with sulfate and oxalate ligands, displaying saturated first coordination spheres (by the coordinated ligands, in fact, always in bidentate modes, and by the added water molecules) are shown in Figure~\ref{fig:fcs-6structures}. This saturation was confirmed by calculations with the addition of one additional water molecule, which always ended up to structures for which one of the explicitly treated water molecules was hydrogen-bonded to one of the anionic ligands. Therefore, we conclude that the coordination number (CN) of the Pa(V) ion can be either 8 ($i$ = 1, 2) or 7 ($i$ = 3). For the \ce{[PaO(C2O4)3]^{3-}} complex, our CN notably matches the EXAFS structure\cite{mendes2010thermodynamical}.

It is clear from Figure~\ref{fig:fcs-6structures} that for each reaction, both the complexes display similar interactions. Thus, we can readily proceed with the determination of the reaction constants (see ``PCM + $n$ \ce{H2O}'' / "FCS'' column in Table~\ref{tab:constants}). For the sake of pedagogy, we will now discuss the results in reversed order, \textit{i.e.} from R3 till R1. Since there is no added water molecule in the \ce{[PaOX3]^{3-}} complexes, note that the value $\text{log}~K_{exc,3}$ of 16.1 is simply reported from the (previous) ``PCM'' column. 

Things start to be less trivial with the \ce{[PaOX2]^{-}} complexes. In both the sulfate and oxalate complexes, three water molecules are introduced. As before, we have checked that each series of three structures, namely GP(DFT/PBE0), GP(HF) and PCM(HF) qualitatively match, meaning that these correspond to a same isomer and conformer. Also, we have excluded the possibility for the need of considering an ensemble of conformers, since no other conformer was found at less than \SI{10}{\kjoule} above the retained one in terms of Gibbs free energy (both in the gas phase and after introducing the PCM correction). We now see a significant improvement of the computed $\text{log}~K_{exc,2}$ value, since it is now for the first time below the $\text{log}~K_{exc,3}$ one. In fact, the $\text{log}~K_{exc,3}-\text{log}~K_{exc,2}$ value is already in good agreement with the experiment (4.1 \textit{vs.} 3.0). Therefore, we conclude that the improvement of our computational description goes together with a better accord with the experiment, which is quite encouraging for what is to follow. 

In the case of R1, we have not reported any value in Table~\ref{tab:constants}. In fact, depending on the starting conformer, we have obtained $\text{log}~K_{exc,1}$ values ranging from 10.0 to -11.6, and we have been unable to find couples of GP(HF) and PCM(HF) structures that match. In this particular case, the static approach combined with a PCM correction seems to act as a random number generator and it is not recommended to report any data. We thus need to figure out how to obtain more stable value in the R1 case, which may come with a change in the chemical process (more water molecules, change in chemical composition) or even of paradigm (by going beyond the static approach for instance). The latter possibility being out of the scope of the present paper, we will continue by proceeding with the other two possibilities. 

Despite the relative favorable correspondence with the experimental data for R2 and R3, it is still important to check that a potential change in the process does not destroy this mighty agreement. In particular, it is important to check that addition of one water molecule outside the first coordination shell (denoted FCS+1 in the remainder of the article), which may physically lead to a better stabilization of one of the dianionic ligands, maintains this successful trend. 

\subsubsection{Beyond the first coordination sphere (FCS+1)}

With the addition of one more water molecule in each complex, an interaction with one of the ligands is actually further introduced. As before, we find a pathological behaviour in the R1 case, and regular behaviour in the R2 and R3 cases. As can be seen in Table~\ref{tab:constants}, the computed $\text{log}~K_{exc,2}$ and $\text{log}~K_{exc,3}$ values become practically closer to the experimental ones while the $\text{log}~K_{exc,3}-\text{log}~K_{exc,2}$ difference remains in fair agreement with experiment. We thus conclude that the previous encouraging results obtained for R2 and R3 were not fortuitous, since improvement of the description maintains them. One should point out that using FCS+2 configurations did not bring any substantial changes to the computed  $\text{log}~K_{exc,n}$ values, further confirming the protactinium(V) coordination number and the thermodynamic stability of considered species. Moreover, since addition of one or two water molecules does not solve the pathological situation encountered in the R1 case, we pursue by increasing the chemical complexity. Since we have seen that the FCS+1 results are overall better than the FCS ones, we stick to the FCS+1 level in this complementary study.  

\subsection{Probing the relevance of the hydroxyl-group scenario}

All the previous results were based on the hypothesis that the hydroxyl group, present in the \ce{[PaO(OH)]^{2+}} starting species, is withdrawn by any complexation. While this is quite clear to us in the R3 case (see the Choice of the systems and working hypotheses section), for which we have obtained a good agreement with the experiment, this hypothesis may be questioned in the R1 and R2 cases, in particular the R1 one for which we were so far unable to report any reliable value. If we make the opposite hypothesis, \textit{i.e.} that the hydroxyl group is maintained, the global formation reactions of the complexes of interest now write:
\begin{equation}
      \ce{[PaO(OH)]^{2+} + i X^{2-} <=>[$\beta_{i}^\prime$] [PaO(OH)X_i]^{2-2i}}
\end{equation}
 \noindent where we recall $i$ is either 1 or 2. These global formation constants translate into the following ligand-exchange reactions:
\begin{equation}\label{eq:reaction_oh}
 \begin{split}
 \ce{[PaO(OH)(SO_4)_i(H2O)_n]^{2-2i} + i C_2O_4^{2-} <=>[$K_{exc,i}^\prime$]} \\ \ce{[PaO(OH)(C_2O_4)_i(H2O)_n]^{2-2i} + i SO_4^{2-}}    
 \end{split}
 \end{equation}
 \noindent To distinguish these reactions from the previous ones (referred to as R1 and R2), they will be denoted as R1-OH and R2-OH, respectively, in the rest of the article. 
 
\begin{figure*}[!htbp]
 \centering
 	\begin{minipage}[t]{0.24\linewidth}	
		\includegraphics[height=2.4cm]{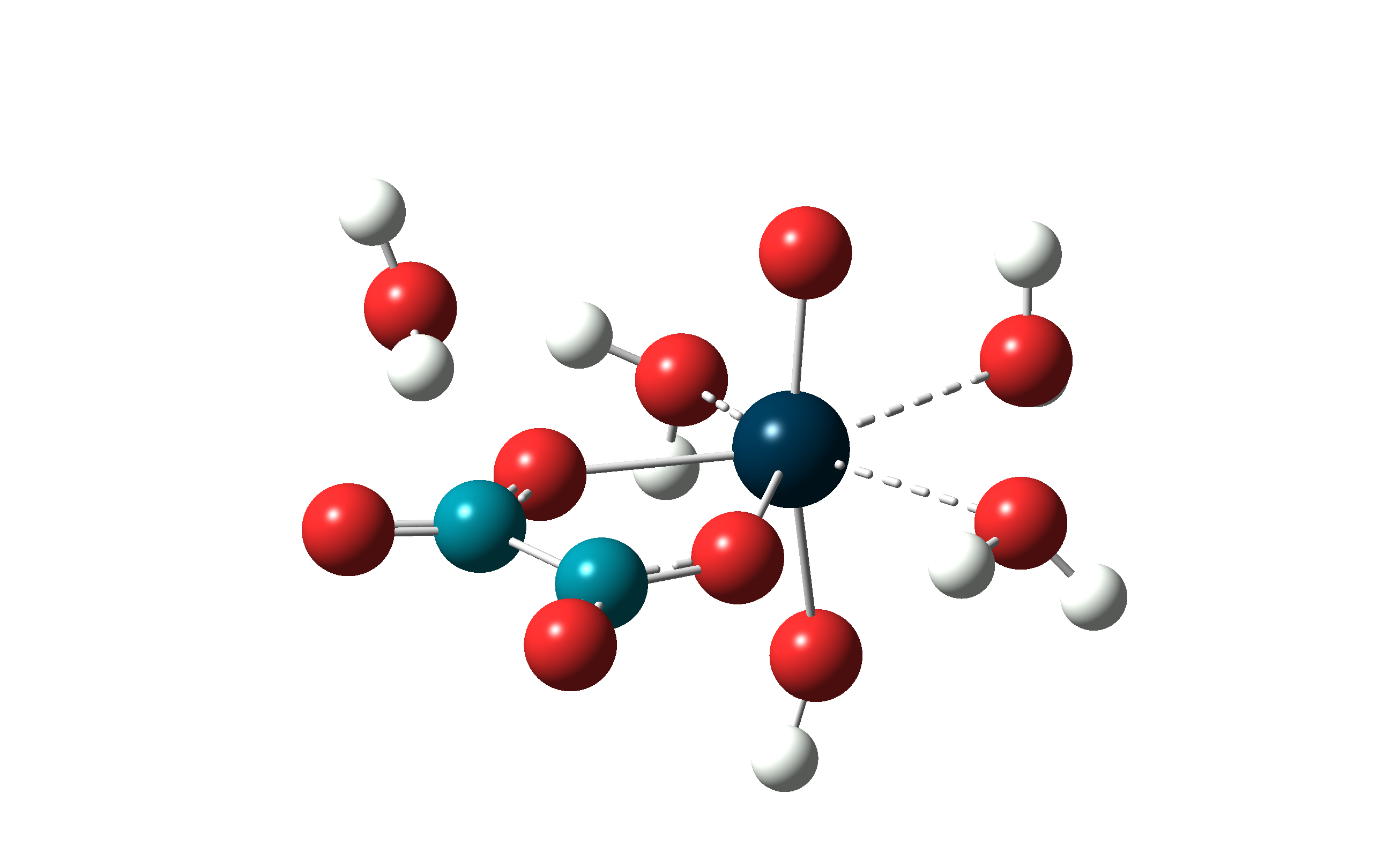}
		\subcaption{\ce{[PaO(OH)(C2O4)(H2O)3]},~\ce{H2O}}
	\end{minipage}
	\begin{minipage}[t]{0.24\linewidth}
		\includegraphics[height=2.4cm]{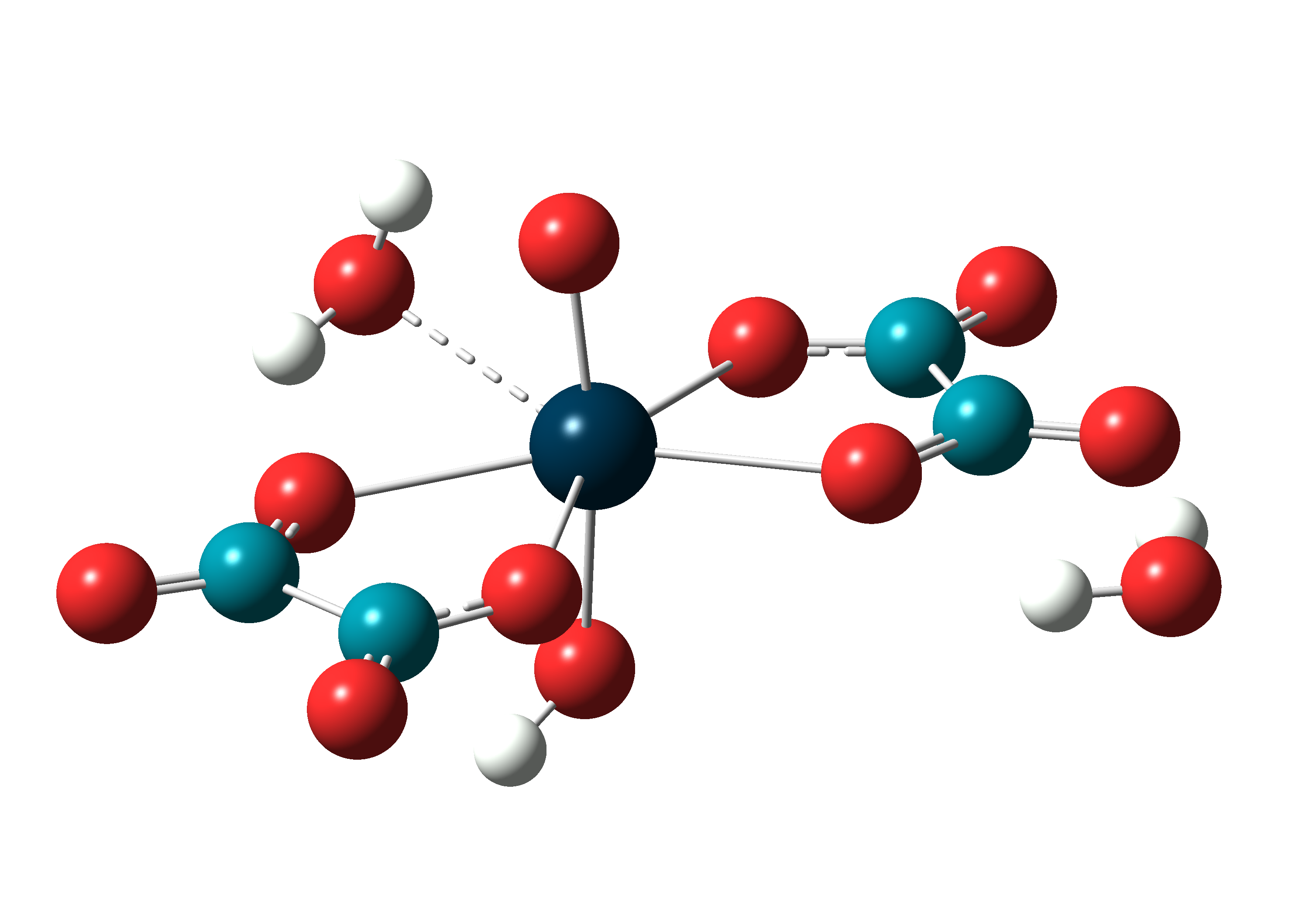}
		\subcaption{\ce{[PaO(OH)(C2O4)2(H2O)]^{2-}},~\ce{H2O}}
	\end{minipage}
	\begin{minipage}[t]{0.24\linewidth}
		\includegraphics[height=2.4cm]{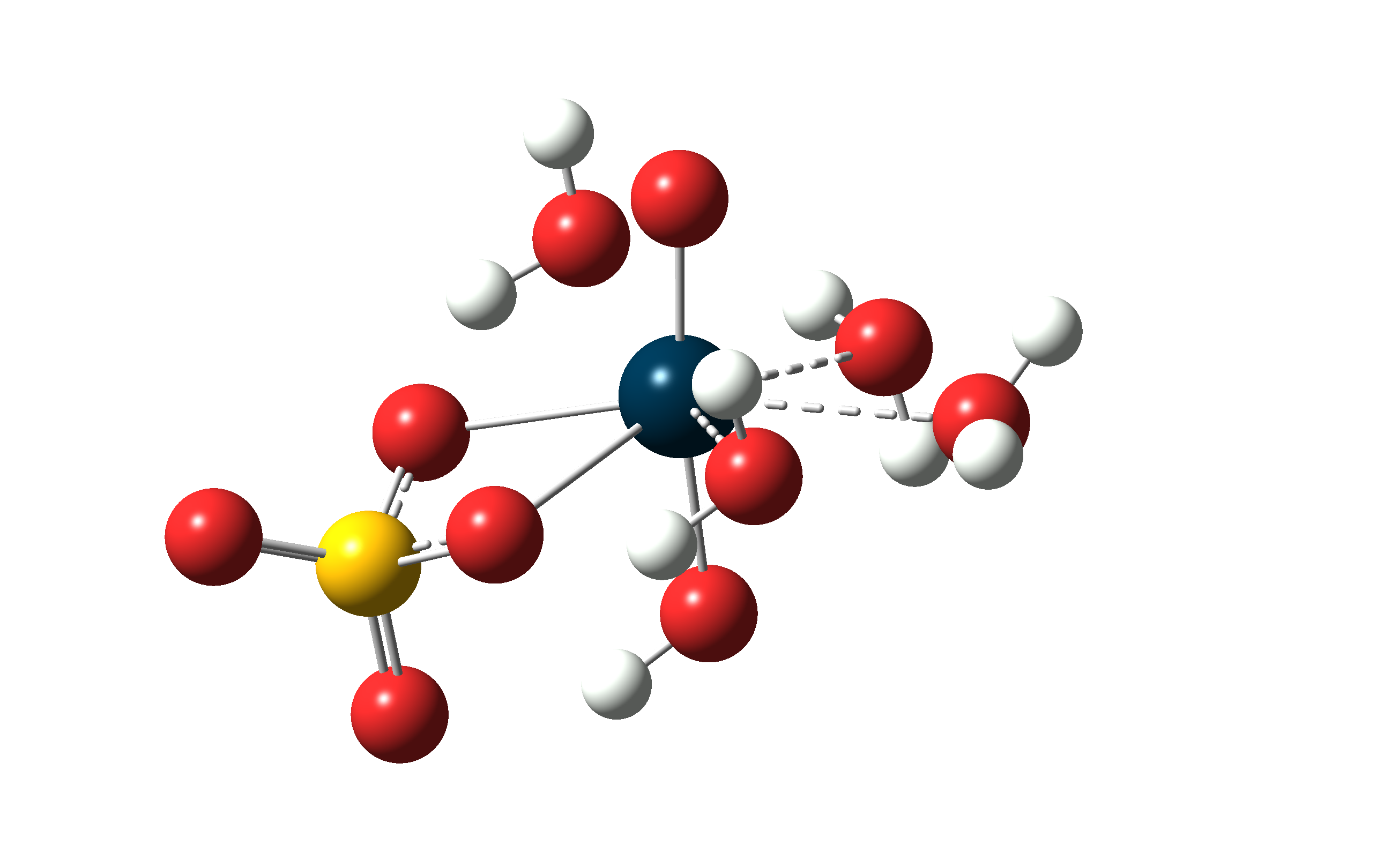}
		\subcaption{\ce{[PaO(OH)(SO4)(H2O)3]},~\ce{H2O}}
	\end{minipage}
	\begin{minipage}[t]{0.24\linewidth}	
	\includegraphics[height=2.4cm]{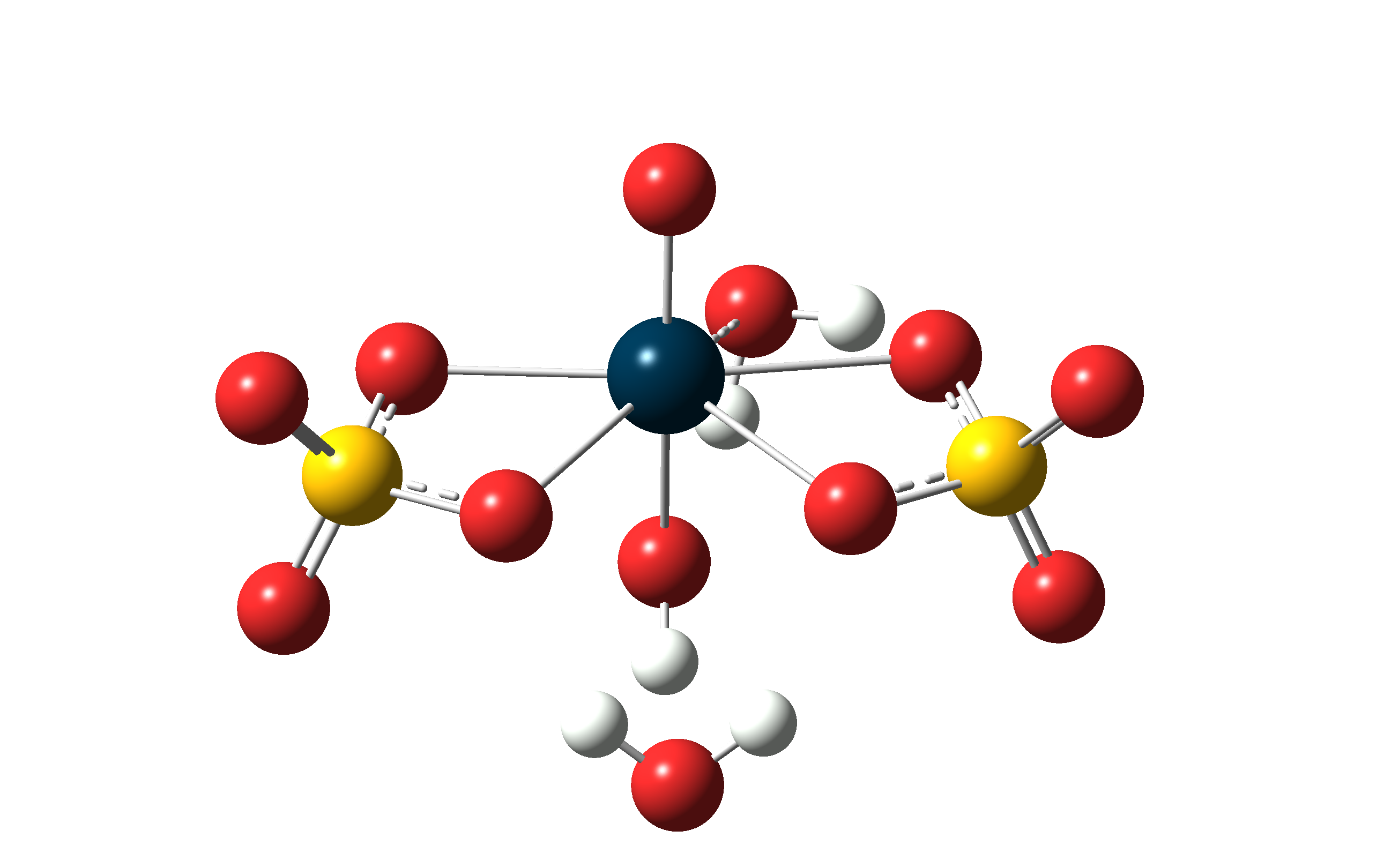}
	\subcaption{\ce{[PaO(OH)(SO4)2(H2O)]^{2-}},~\ce{H2O}}
\end{minipage}
 \caption{Representations of the GP(DFT/PBE0) structures of the complexes involved in the R1-OH and R2-OH reactions, obtained with one more water than just saturated first-coordination spheres (FCS+1). Color code: dark blue (Pa), yellow (S), red (O), light blue (C) and white (H).}
 \label{fig:hydroxyl-str}
\end{figure*}

 The four new FCS+1 GP(DFT/PBE0) structures are displayed in Figure~\ref{fig:hydroxyl-str}. The hydroxyl group is always found to be in \textit{trans} position with respect to the mono-oxo one, and the sulfate and oxalate ligands are still bound in the bidentate fashion, as in all the previous structures. In those four new structures, the CN of the Pa(V) ion is 7. Luckily, this reduces the possibilities for arranging the water molecules within the FCS and also further away from it. In practice, we now have the expected correspondence between the GP(DFT/PBE0), GP(HF) and PCM(HF) structures even in the complexes that are involved in R1--OH, meaning that we may now propose properly determined log $K_{exc,i}$ values for all reactions. Also, we now have another log $K_{exc,2}$ value, associated with the R2-OH equilibrium. Those two new values, together with our previous best estimates (FCS+1) are reported in Table~\ref{tab:final}.  
 
\begin{table}[h]
\small
  \caption{\ Computed values of the logarithms of the ligand-exchange reaction constants as functions of the chemical reaction (R/R-OH). Calculations are performed with one more water than just-saturated first-coordination spheres (FCS+1). Values are repeated from Tables \ref{tab:log_beta_exp} and \ref{tab:constants} for convenience.}
  \label{tab:final}
  \begin{tabular*}{0.48\textwidth}{@{\extracolsep{\fill}}lll}
    \hline
    log $K_{exc,i}/K_{exc,i}^\prime$ & Comput. & Expt.  \\
    \hline
    R1/R1-OH & - / 7.0& 3.5  \\
    R2/R2-OH & 8.8 / 12.1 & 7.5 \\
    R3/R3-OH & 14.6 / - & 10.5 \\
    \hline
  \end{tabular*}
\end{table}

As can be seen in this Table, the values obtained for R1-OH and R2-OH are both quite consistent with the experiment. Moreover, the $\text{log}~K_{exc,2}-\text{log}~K_{exc,1}$ difference is also in good agreement with the experiment. Even if this agreement, based on the analysis of apparent constants for the experimental values, may not be a definitive proof for the occurrence of the hydroxyl-group scenario in the R1 case (more experimental information may be required), we conclude by predicting that this scenario must be at play since our static approach is found to fail at reproducing the trends otherwise. 

Concerning the R2 reaction, both scenarios lead to computed $\text{log}~K_{exc,2}$ values in decent accord with the experiment. Therefore, it is hard to know \textit{a priori} if the hydroxyl group is maintained or withdrawn in the 1:2 complexation. Moreover, being in fact an intermediate situation sandwiched between the two clear R1 and R3 cases, it could happen that the hydroxyl group is maintained in the sulfate complex and not in the oxalate one. Unfortunately, this cannot be well probed with our calculations since we need to have similar pictures in both the left-hand and right-hand sides of the reactions to favor error cancellations, especially on the solvation contributions.

\section{Concluding remarks}
In this work, we performed a step-by-step study to deduce ligand-exchange reaction constants. The studied equilibria involved Pa(V) complexes with sulfate and oxalate ligands, displaying the characteristic Pa mono-oxo bond, and in some cases hydroxyl groups. Our computational approach is based on the DFT/PBE0 level to determine the gas-phase complexation constants, and on the UAHF model for the solvation corrections. In the absence of an alternative, note that the UFF radius has been used for protactinium.

We have seen that solvation of Pa(V) complexes may be tricky, in the sense that the implicit model alone does not suffice to reach good values, even if relative constants are looked for. Instead, we show that we need to at least saturate the coordination sphere of the Pa(V) ion by water addition. Also, in the case of low-coordinated complexes, we reveal that the speciation may be more complex than expected and that the relevant complexes may still display one hydroxyl group.

The determination of the speciation of Pa(V) in non-complexing and in complexing media is still in its infancy. It is extremely hard to firmly identify new chemical species, and we believe that our work, which has led to the validation of a computational protocol, can open the way for future studies assisted by computations. In particular, we have shown that calculations of ligand-exchange reaction constants can be quite informative concerning trends (which ligands is a stronger complexant) and also regarding the occurrence of given chemical groups and/or bonds in the formed complexes. 

\section*{Author Contributions}

H.O. and R.M. defined the working hypotheses and the computational strategy. H.O. was responsible for performing and/or checking all the calculations. J.D. performed preliminary calculations concerning the hydroxyl-group scenario. The first draft of the manuscript was written by H.O. and R.M. All the authors contributed to the scientific development of the work and were given the opportunity to amend the manuscript.

\section*{Conflicts of interest}

The authors declare no conflict of interest.

\section*{Acknowledgements}

This work was supported by the \textit{Région des Pays de la Loire} (RCT-PaPAn project) and by the French National Research Agency (ANR-21-CE29-0027, LABEX CaPPA/ANR-11-LABX-0005-01 and I-SITE ULNE/ANR-16-IDEX-0004 ULNE). HPC resources from CCIPL (\textit{Centre de Calcul Intensif des Pays de la Loire}) were used.

\bibliography{protactinium} 
\bibliographystyle{rsc} 

\clearpage



\section*{Supplementary material}

# Supporting Information:

# Coordination and thermodynamic properties of aqueous protactinium(V) by first-principle calculations

Hanna Oher,[†,‡] Jeremy Delafoulhouze,[¶] Eric Renault,[¶] Valérie Vallet,[§] and Rémi Maurice[*,†,‡]

†SUBATECH, UMR CNRS 6457, IN2P3/IMT Atlantique/Université de Nantes, 4 rue Alfred Kastler, BP 20722, 44307 Nantes Cedex 3, France

‡Univ Rennes, CNRS, ISCR (Institut des Sciences Chimiques de Rennes) – UMR 6226, 35000 Rennes, France

¶Nantes Université, CNRS, CEISAM UMR 6230, F-44000 Nantes, France

§Univ. Lille, CNRS, UMR 8523-PhLAM-Physique des Lasers, Atomes et Molécules, F-59000 Lille, France

E-mail: remi.maurice@univ-rennes.fr

# Preparatory study on molecular geometries

Not many theoretical studies have been dedicated to protactinium(V) complexes. Therefore, we have performed a preparatory study on molecular geometries, with special emphasis on the $[\text{PaO}(\text{C}_2\text{O}_4)_3]^{3-}$ complex, for which both an EXAFS structure and a previously optimized B3LYP one were reported.[S1] We focus on key bond distances. In particular, we have tested:

1. Three electronic structure methods: MP2,[S2] DFT/B3LYP[S3] and DFT/PBE0;[S4,S5]

2. Two schemes for introducing the scalar relativistic effects, that is via the use of the all-electron Douglas-Kroll-Hess Hamiltonian[S6–S8] or via the use of a scalar relativistic energy-consistent pseudopotential, namely ECP60MWB;[S9]

3. Three basis sets, one all-electron one (SARC-DKH2[S10] for Pa and def2-TZVP[S11] for the other atoms) and two that are compatible with the ECP60MWB pseudopotential (with a segmented basis set[S9] for Pa and the def2-TZVP[S11] or aug-cc-VTZ[S12] ones for the other atoms);

4. Three different quantum chemistry codes, namely Gaussian,[S13] ORCA 4.2.1[S14] and OpenMolcas.[S15]

Not all the combinations have been tested, since a more limited number of cases was sufficient to reach the key conclusions that we needed to further proceed with the computations of ligand-exchange reaction constants. Results are presented in Tables S1 and S2.

The main objective of Table S1 was to compare results of all-electron calculations (columns 4 and 5) and of pseudopotential-based ones (columns 2 and 3). Three codes were used to perform these tests, because specific implementations that are not available in all the codes were required. We were first puzzled by the ORCA/B3LYP all-electron result, largely differing from the Gaussian/B3LYP one for which a pseudopotential was used. We have thus performed additional calculations with OpenMOLCAS. First, we have repeated the Gaussian/B3LYP calculation with the same generic setup (ECP60MWB/def2-TZVP), obtaining

Table S1: Comparison of codes and of all-electron *vs.* pseudopotential-based calculations for determining the structure of the $[PaO(C_2O_4)_3]^{3-}$ complex in the gas phase (all the distances are in Å).

|  | Gauss./B3LYP ECP60MWB def2-TZVP | OpenM./B3LYP ECP60MWB def2-TZVP | ORCA/B3LYP SARC-DKH2 def2-TZVP | OpenM./B3LYP SARC-DKH2 def2-TZVP |
|---|---|---|---|---|
| Pa–O$_{oxo}$ | 1.862 | 1.862 | 1.792 | 1.871 |
| av. Pa–O$_{ligand}$ | 2.386 | 2.386 | 2.354 | 2.390 |
| av. Pa–C | 3.291 | 3.291 | 3.253 | 3.295 |
| av. Pa–O$_{ligand}'$ | 4.477 | 4.447 | 4.440 | 4.481 |

the exact same result. Thus, both the Gaussian and OpenMOLCAS implementations for pseudopotential calculations agree.

Then, we have repeated the all-electron SARC-DKH2/def2-TZVP calculations, finding a very different outcome than what was found with ORCA. After noticing the close matching between the all-electron OpenMOLCAS result and both the pseudopotential-based ones, we have looked for potential sources of errors in the ORCA/B3LYP calculation. It appeared that an extra simplification, in particular a one-center approximation, was present in the ORCA implementation for geometry optimization with the DKH2 Hamiltonian. Consequently, the ORCA/B3LYP result can be discarded and we conclude that the molecular geometries can be safely computed with both the ECP60MWB pseudopotential or with all-electron basis sets. For the sake of efficiency, we have thus chosen to follow the pseudopotential approach.

Table S2: Comparison of the B3LYP, PBE0 and MP2 optimized structures of the $[PaO(C_2O_4)_3]^{3-}$ complex in the gas phase (all distances in Å). All the calculations have been performed with Gaussian and made use of the ECP60MWB pseudopotential.

|  | EXAFS[S1] | B3LYP[S1] 6-31+G* | B3LYP def2-TZVP | PBE0 def2-TZVP | MP2 def2-TZVP | B3LYP aug-cc-pVTZ | PBE0 aug-cc-pVTZ | MP2 aug-cc-pVTZ |
|---|---|---|---|---|---|---|---|---|
| Pa–O$_{oxo}$ | 1.75 | 1.89 | 1.862 | 1.843 | 1.855 | 1.863 | 1.845 | 1.862 |
| av. Pa–O$_{ligand}$ | 2.38 | 2.39 | 2.386 | 2.364 | 2.353 | 2.387 | 2.365 | 2.347 |
| av. Pa–C | 3.28 | 3.30 | 3.291 | 3.265 | 3.247 | 3.293 | 3.267 | 3.244 |
| av. Pa–O$_{ligand}'$ | 4.47 | 4.50 | 4.477 | 4.447 | 4.439 | 4.480 | 4.451 | 4.435 |

In Table S2, we have reported the previous EXAFS and B3LYP results,[S1] together with

new calculations with more advanced basis sets, namely the def2-TZVP and aug-cc-pVTZ ones. A few points deserve to be commented:

1. When all the B3LYP results are compared, two main behaviours are found: with the 6-31+G* basis set, all the bond distances of interest seem overestimated, while a more consistent behaviour is observed with both the def2-TZVP and aug-cc-pVTZ basis sets, which are thus recommended;

2. For the three tested methods, the results obtained with either the def2-TZVP or aug-cc-pVTZ basis set are of equal quality; the def2-TZVP basis set, which is smaller, is thus preferable for efficiency purposes;

3. The MP2, DFT/PBE0 and DFT/B3LYP bond distances are very similar, all in good accord with the EXAFS structure, despite a remaining discrepancy on the Pa–$O_{oxo}$ bond;

4. The DFT/PBE0 method gives the Pa–$O_{oxo}$ bond distance in the closest agreement with experiment.

All in all, we have chosen to use the DFT/PBE0 method, the small-core ECP60MWB pseudopotential for Pa, and the def2-TZVP basis set for all the light atoms.

# Supplemental figure

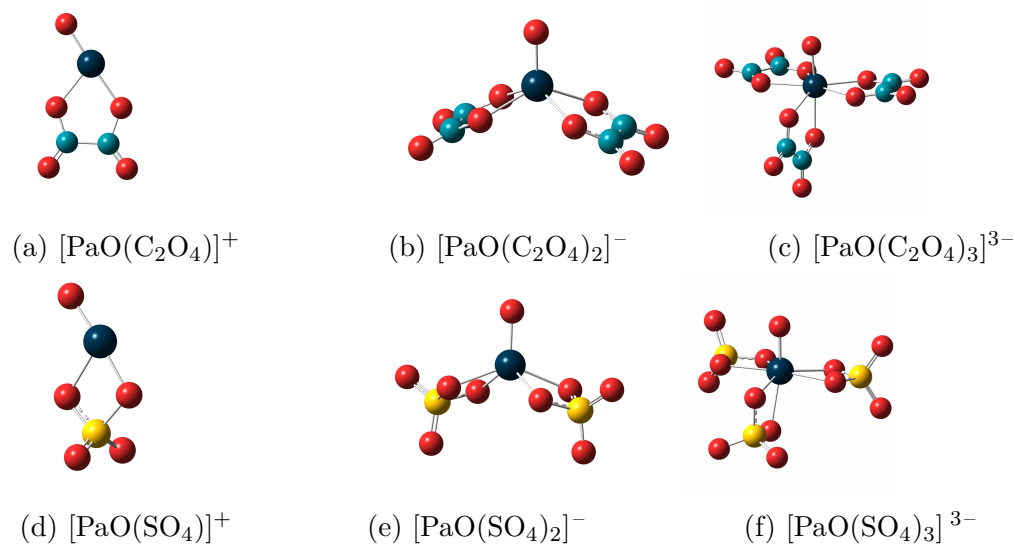

(a) [PaO(C$_2$O$_4$)]$^+$  (b) [PaO(C$_2$O$_4$)$_2$]$^-$  (c) [PaO(C$_2$O$_4$)$_3$]$^{3-}$

(d) [PaO(SO$_4$)]$^+$  (e) [PaO(SO$_4$)$_2$]$^-$  (f) [PaO(SO$_4$)$_3$]$^{3-}$

Figure S1: Representations of the GP(DFT/PBE0) structures of the Pa(V) complexes, obtained with unsaturated first-coordination spheres (corresponding to the structures of the bare complexes used in the "GP" and "PCM" columns of Table 2). Color code: dark blue (Pa), yellow (S), red (O) and light blue (C).



\end{document}